%% file: main.tex
\documentclass[conference]{IEEEtran}
\IEEEoverridecommandlockouts
\usepackage{cite}
\usepackage{amsmath,amssymb,amsfonts}
\usepackage{graphicx}
\usepackage{textcomp}
\usepackage[dvipsnames]{xcolor}
\usepackage{booktabs}
\usepackage{multirow}
\usepackage{array}
\usepackage{mdframed}
\usepackage{adjustbox}
\usepackage{listings}
\usepackage{enumitem}
\usepackage{fontawesome5}
\usepackage{subcaption}

\usepackage{tikz}
\usetikzlibrary{arrows.meta, shapes}
\usetikzlibrary {shadows}
\usetikzlibrary{fit, backgrounds, calc, positioning}
\usetikzlibrary{fadings}
\usetikzlibrary{decorations.pathmorphing}

\usepackage{algorithm}
\usepackage{algorithmicx}
\usepackage[noend]{algpseudocode}

\usepackage{xurl}
\usepackage{hyperref}
\usepackage{cleveref}

\algrenewcommand\algorithmiccomment[2][\itshape]{{#1\hfill\(\triangleright\)
    #2}}
\algrenewcommand{\algorithmicrequire}{\textbf{Input:}}
\algrenewcommand{\algorithmicensure}{\textbf{Output:}}

\newmdenv[
  backgroundcolor=gray!10,
  linecolor=black,
  linewidth=1pt,
  roundcorner=4pt,
  innertopmargin=6pt,
  innerbottommargin=8pt,
  innerleftmargin=8pt,
  innerrightmargin=8pt,
  skipabove=12pt,        
  skipbelow=12pt,        
  frametitlebackgroundcolor=black,
  frametitlefont=\bfseries\color{white},
  frametitlerule=false,
  frametitlealignment=\raggedright,
  frametitleaboveskip=6pt,
  frametitlebelowskip=6pt,
]{findingbox}

\newcommand{\inlineheadingbf}[1]{\medskip\noindent{\bfseries #1.}}
\newcommand{\inlineheadingit}[1]{\medskip\noindent{\it #1.}}

\def\BibTeX{{\rm B\kern-.05em{\sc i\kern-.025em b}\kern-.08em
    T\kern-.1667em\lower.7ex\hbox{E}\kern-.125emX}}

\makeatletter
\newcommand{\linebreakand}{%
  \end{@IEEEauthorhalign}
  \hfill\mbox{}\par
  \mbox{}\hfill
  \begin{@IEEEauthorhalign}
}
\makeatother

\begin{document}

\title{Prompt Coverage Adequacy}

\author{\IEEEauthorblockN{Florian Tambon}
\IEEEauthorblockA{
\textit{SnT, University of Luxembourg}\\
Luxembourg \\
florian.tambon@uni.lu}
\and
\IEEEauthorblockN{Michael Konstantinou}
\IEEEauthorblockA{
\textit{SnT, University of Luxembourg}\\
Luxembourg \\
michael.konstantinou@uni.lu}
\and
\IEEEauthorblockN{Cedric Richter}
\IEEEauthorblockA{
\textit{SnT, University of Luxembourg}\\
Luxembourg \\
cedric.richter@uni.lu}
\linebreakand
\IEEEauthorblockN{Charles Chenouard}
\IEEEauthorblockA{
\textit{SnT, University of Luxembourg}\\
Luxembourg \\
charles.chenouard@ext.uni.lu}
\and
\IEEEauthorblockN{Mark Harman}
\IEEEauthorblockA{
\textit{University College London}\\
London, UK\\
mark.harman@ucl.ac.uk}
\and
\IEEEauthorblockN{Mike Papadakis}
\IEEEauthorblockA{
\textit{SnT, University of Luxembourg}\\
Luxembourg \\
michail.papadakis@uni.lu}
}

\maketitle

\begin{abstract}
In recent years, it has become increasingly evident that large language models (LLMs) and autonomous agents raise the level of abstraction in software development by shifting the focus from writing precise procedures to expressing intents and goals. This paradigm shift introduces new challenges, particularly in how testing should be guided when prompts, rather than code, become primary development artifacts. To address this challenge, we propose Prompt Coverage Adequacy, a novel coverage criterion designed to support the testing of code generated from task descriptions. Prompt Coverage Adequacy serves as an analog to traditional code coverage, but operates at the level of prompts used in LLM and agent-based programming. Specifically, it measures how well a given test suite satisfies the requirements expressed in a prompt by leveraging the attention mechanisms of LLMs. We evaluate a simple instantiation of this criterion, based on attention boosting, across two datasets and multiple LLMs. Our results demonstrate that Prompt Coverage is associated with fault-detection effectiveness and can uncover over 30+\% more faults than traditional code coverage when used to guide test generation. These findings suggest that Prompt Coverage Adequacy can serve as a foundation for developing testing metrics better suited to the emerging paradigm of LLM-driven software development, addressing the limitations of classical coverage criteria in this new context.
\end{abstract}

\begin{IEEEkeywords}
language models, requirements, testing, coverage criteria 
\end{IEEEkeywords}

\input{sections/intro}
\input{sections/background}

\input{sections/method}

\input{sections/evaluation}
\input{sections/results}
\input{sections/threats}

\input{sections/related}
\input{sections/conclusion}

\IEEEtriggeratref{36} 
\bibliographystyle{IEEEtranS}
\bibliography{literature}
\end{document}

%% file: sections/intro.tex
\section{Introduction}

Large Language Models (LLMs) can automatically generate software from natural language descriptions, significantly accelerating development. However, the generated code may contain bugs, misunderstand requirements, omit functionality, or introduce security vulnerabilities. Since LLMs do not truly understand the intended behavior of a system, they can produce code that appears correct but fails in important situations \cite{TambonDNKDA25}.

Traditionally, software testing relies on adequacy criteria \cite{ZhuHM97,PapadakisK00TH19}, such as statement, branch, and mutation testing, to assess the effectiveness of a test suite. These criteria provide measurable indicators of test strength, help identify untested functionality, guide the creation of additional tests, and support the prioritization of testing effort. More broadly, test adequacy criteria offer an objective basis for evaluating how thoroughly a software system has been exercised.

However, the emergence of LLM-driven software development challenges the assumptions underlying traditional coverage metrics. In conventional software engineering, source code constitutes the primary development artifact, and coverage criteria evaluate the extent to which the generated code has been executed by a test suite. In contrast, when software is produced from natural language task descriptions, prompts become first class artifacts that explicitly capture the intended requirements and behavior of the system \cite{fawzy2025vibecodingpracticemotivations}.

This shift exposes a fundamental limitation of traditional coverage metrics. Consider a prompt that requests a booking system supporting booking creation, cancellation, refunds, and email notifications. A test suite may exercise all generated code and achieve near complete or even 100\% code coverage while validating only booking creation and cancellation. In such a scenario, critical requirements such as refunds and notifications may remain entirely untested. While traditional coverage metrics can reveal which portions of the generated code have been executed, they provide little insight into whether all requirements expressed in the original prompt have been adequately validated.

\textit{As software development becomes increasingly prompt-driven, testing methodologies must evolve beyond measuring code execution alone and explicitly assess the extent to which prompt specified requirements are exercised by tests}. This observation motivates the need for novel adequacy criteria that operate at the level of user intent rather than solely at the level of implementation. In particular, Prompt Coverage Adequacy aims to quantify how comprehensively a test suite validates the requirements expressed in a prompt, providing a complementary view to traditional code coverage. Such criteria have the potential to establish a measurable notion of completeness for testing LLM generated software and to improve confidence on its correctness.

Beyond identifying inadequately tested prompt requirements, prompt level adequacy criteria offer several benefits for the development and assurance of LLM generated software. 
\begin{itemize}
    \item First, they provide \textit{improved traceability between requirements and tests }by explicitly linking portions of a prompt to the test cases that validate them. This allows developers to reason not only about which code has been executed, but also about which user requirements have been exercised and which remain unverified.
    \item Second, prompt based adequacy metrics can support the \textit{systematic generation of tests}. By identifying prompt requirements that exhibit low coverage, developers or automated testing tools can focus test creation efforts on under tested aspects of the specification. Such guidance is particularly valuable in LLM driven development, where generated implementations may contain functionality that is difficult to interpret from the source code alone.
    \item Third, these criteria can improve the \textit{evaluation and comparison of generated solutions}. Two implementations may achieve similar traditional coverage scores while exhibiting substantially different levels of compliance with the original prompt. Prompt coverage provides an additional dimension for assessing whether generated software faithfully realizes the requested functionality.
    \item Fourth, prompt level adequacy criteria can \textit{facilitate the detection of common failure modes of code} including omitted requirements, partially implemented features, and incorrect interpretations of user intent. Since these defects originate at the specification level, they may remain undetected by conventional code centric metrics even when all generated code is exercised during testing.
\end{itemize}

Additionally, prompt level adequacy criteria provide a natural mechanism for human and AI communication and accountability. By making explicit which prompt requirements have been validated, these criteria offer developers, reviewers, and stakeholders greater visibility into the relationship between intent, implementation, and testing. Such transparency can increase confidence in generated software and support certification, auditing, and quality assurance processes.

More broadly, prompt coverage represents a shift from measuring the completeness of testing with respect to an implementation toward measuring completeness with respect to the natural language specification. As prompts become first class software artifacts in LLM-based development, adequacy criteria defined at the prompt level may become as fundamental to quality assurance as code coverage is in traditional software engineering.

In technical terms, we introduce the prompt coverage criterion based on the attention mechanisms of large language models. Specifically, we instruct the model to perform a consistency check between candidate test cases and their related task descriptions (prompts). We then quantify the resulting change in the model’s output entropy when attention is selectively increased (or spotlighted~\cite{DBLP:conf/eacl/VenkateswaranC26}) toward candidate natural language test requirements. This entropy variation serves as an indicator of the extent to which the test requirements are covered by, and aligned with, the prompt.

We evaluate the proposed criterion on two widely used benchmarks, HumanEval~\cite{DBLP:journals/corr/abs-2107-03374} and LiveCodeBench~\cite{DBLP:conf/iclr/JainHGLYZWSSS25}, and across two large language models, Qwen2.5 and Gemma3. Our results demonstrate that prompt coverage is a meaningful testing metric, exhibiting a strong correlation with traditional code coverage. Specifically, higher prompt coverage consistently leads to higher code coverage, despite not relying on any information about the underlying implementation. Furthermore, we show that increasing prompt coverage improves fault detection effectiveness independently of test suite size, indicating that the increased test effectiveness is derived from the properties of prompt coverage rather than simply from generating more tests. 

Perhaps more importantly, our experiments show that coverage-driven test generation based on prompt coverage achieves approximately 30\% higher fault detection than generation guided by code coverage. In particular, Prompt Coverage-driven tests reveals 49\% and 57\% of the faults in HumanEval and LiveCodeBench, while Code Coverage-driven tests reveal 18\% and 24\%. These findings suggest that prompt coverage offers a direct link between natural language specifications and expected (tested) program behaviour, forming an objective way that guides LLM-based test generation. 

Overall, we believe that Prompt Coverage Adequacy can provide the foundation for a new class of testing criteria/metrics tailored to the emerging paradigm of LLM-driven software development. LLM-generated software originates from natural-language specifications, where the prompt serves as the initial and often most influential description of the system's intended behavior.

Furthermore, since Prompt Coverage shifts the focus from the code to its specifications, it remains robust during multiple rounds of LLM-based development, where generated code may significantly vary in structure while still satisfying the same prompt. In these cases, the coverage of the code would be less reliable as a proxy for behavioral adequacy. Similarly, by evaluating whether tests exercise different aspects of the prompt, Prompt Coverage is applicable even when the implementation is unavailable or incomplete.


As software development increasingly relies on LLMs testing methodologies must evolve accordingly. Prompt Coverage represents a first step towards natural language specification-aware adequacy criteria that evaluate test suites based on their ability to cover the semantic space of user requirements rather than solely the structural characteristics of a particular implementation. Such metrics may ultimately enable more effective testing, validation, and quality assurance workflows for generated software, where understanding and exercising the prompt can be as important as exercising the code. We make our replication package available \cite{ReplicationPackage}

\begin{figure}[t]
\vspace{0.5em}
\centering
\resizebox{\linewidth}{!}{\input{figures/motivation}}
\vspace{0.5em}
\caption{\textit{Prompt Coverage Adequacy.} Code generation prompt of HumanEval/103 and its corresponding test suite. Arrows indicate coverage relations computed with Qwen2.5-Coder. Both test case 1 and 2 cover the behavior detailed by {\color{teal!80!black}requirement 1} and {\color{RoyalBlue!70!black}requirement 2} (first and second sentence), while test 3 uniquely covers {\color{BurntOrange}requirement 3}. Without test 3, our criterion would flag {\color{BurntOrange}requirement 3} as uncovered, thereby illustrating aspects of the program behavior that are not adequately exercised by the test cases. This coverage gap can then guide testing towards writing the missing test.}
\label{fig:motivation}
\vspace{0.5em}
\end{figure}
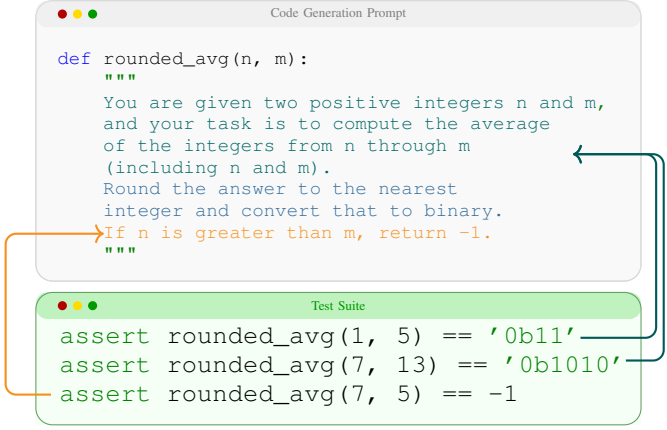

\section{Motivating Example}
In this section, we present the key concept and ideas of prompt coverage using a code-generation prompt as a motivating example. 

\Cref{fig:motivation} shows an example code generation prompt from HumanEval~\cite{DBLP:journals/corr/abs-2107-03374}. HumanEval simulates a code generation scenario where the developer provides a function stub to be implemented, but leaves the implementation to the LLM. 

In this example, the developer instructs the LLM to implement a \texttt{rounded\_avg} function that takes two positive integers \texttt{n} and \texttt{m} and computes the average of the integers between \texttt{n} and \texttt{m}. The docstring defines the following requirements:
\begin{enumerate}[label=\arabic*.]
    \item \texttt{rounded\_avg} should return the average of integers from \texttt{n} through \texttt{m}. 
    \item \texttt{rounded\_avg} should round the answer to the nearest integer and convert it to binary.
    \item \texttt{rounded\_avg} should return $-1$ if \texttt{n} is greater than \texttt{m}.
\end{enumerate}
To validate these requirements, three test cases are provided (out of a larger test suite). Test case 1 and 2 evaluate the regular behavior of the function, while test case 3 checks the edge case (requirement 3). 


We observe that test cases 1 and 2 are associated with requirements 1 and 2, while test case 3 is associated with requirement 3. The key question addressed by our coverage criterion is how such associations can be derived automatically and reliably. To achieve this, we leverage the entropy of large language models as a proxy for the model's level of surprise and employ spotlighting ~\cite{DBLP:conf/eacl/VenkateswaranC26} to direct the model’s attention toward a specific requirement. A requirement is considered covered when focusing the model’s attention on that requirement (spotlighting) leads to a reduction in its surprise, as reflected by a decrease in entropy.

The arrows in \Cref{fig:motivation} show the coverage relations between tests and requirements computed with Qwen2.5-Coder. Requirements 1 and 2 are jointly covered by tests 1 and 2, while requirement 3 is uniquely covered by test 3. Without test 3, our coverage criterion would flag requirement 3 as uncovered, indicating a potential gap in the testing process.

Finally, this example shows the advantages of prompt coverage over more traditional implementation-level coverage criteria. Consider the following {\em faulty} implementation of the prompt specification (as possibly generated by an LLM):
\begin{center}
\vspace{-0.5em}
\adjustbox{max width=0.9\linewidth}{
\begin{lstlisting}[numbers=none, breaklines=false, escapechar=!]
def rounded_avg(n, m):
    summation = 0 
    for i in range(n, m+1): 
        summation += i 
    return bin(round(summation/(m - n + 1)))
\end{lstlisting}
}
\end{center}
Tests 1 and 2 would each achieve 100\% statement coverage for this example, while the test 3 that detects the fault would not cover any additional statements and would be considered as redundant by traditional coverage criteria. Our prompt coverage criterion can therefore highlight testing gaps in behavior that is specified in the prompt but is missing from the code.

%% file: figures/motivation.tex
\begin{tikzpicture}[
    x=1cm,y=1cm,
    every node/.style={inner sep=0pt, outer sep=0pt},
    title/.style={font=\bfseries\fontsize{18}{20}\selectfont},
    sectiontitle/.style={font=\bfseries\fontsize{16}{18}\selectfont},
    bubbletext/.style={font=\ttfamily\fontsize{15}{17}\selectfont, align=center},
    iconlabel/.style={font=\fontsize{22}{24}\selectfont},
]


\node[bubbletext, inner sep=0.2cm, align=left, text width=7.6cm] (prompt) at (0,-1.1) {%
  \adjustbox{max width=0.45\linewidth}{
    \begin{lstlisting}[numbers=none, breaklines=false, escapechar=!]
def rounded_avg(n, m): 
    """
    !\color{teal!80!black}You! !\color{teal!80!black}are! !\color{teal!80!black}given! !\color{teal!80!black}two! !\color{teal!80!black}positive! !\color{teal!80!black}integers! !\color{teal!80!black}n! !\color{teal!80!black}and! !\color{teal!80!black}m!,
    !\color{teal!80!black}and! !\color{teal!80!black}your! !\color{teal!80!black}task! !\color{teal!80!black}is! !\color{teal!80!black}to! !\color{teal!80!black}compute! !\color{teal!80!black}the! !\color{teal!80!black}average! 
    !\color{teal!80!black}of! !\color{teal!80!black}the! !\color{teal!80!black}integers! !\color{teal!80!black}from! !\color{teal!80!black}n! !\color{teal!80!black}through! !\color{teal!80!black}m! 
    !\color{teal!80!black}(including! !\color{teal!80!black}n! !\color{teal!80!black}and! !\color{teal!80!black}m!!\color{teal!80!black}).!
    !\color{RoyalBlue!70!black}Round! !\color{RoyalBlue!70!black}the! !\color{RoyalBlue!70!black}answer! !\color{RoyalBlue!70!black}to! !\color{RoyalBlue!70!black}the! !\color{RoyalBlue!70!black}nearest! 
    !\color{RoyalBlue!70!black}integer! !\color{RoyalBlue!70!black}and! !\color{RoyalBlue!70!black}convert! !\color{RoyalBlue!70!black}that! !\color{RoyalBlue!70!black}to! !\color{RoyalBlue!70!black}binary.! 
    !\color{BurntOrange}If! !\color{BurntOrange}n! !\color{BurntOrange}is! !\color{BurntOrange}greater! !\color{BurntOrange}than! !\color{BurntOrange}m,! !\color{BurntOrange}return! !\color{BurntOrange}-1.! 
    """
    \end{lstlisting}
}};

\node[fill=gray!12, minimum width=8.0cm, minimum height=0.35cm,
      anchor=south, rounded corners=0.15cm] (titlebar) at ([yshift=0cm]prompt.north) {};

\node[text=gray] at (titlebar.center) {\tiny Code Generation Prompt};

\fill[red!70!black]     ([xshift=0.35cm]titlebar.west) circle (0.06cm);
\fill[yellow!80!orange] ([xshift=0.55cm]titlebar.west) circle (0.06cm);
\fill[green!60!black]   ([xshift=0.75cm]titlebar.west) circle (0.06cm);

\begin{scope}[on background layer]
\draw[fill=gray!5, rounded corners=0.22cm, draw=gray!30, line width=0.4pt]
  (titlebar.north west) rectangle (prompt.south east);
\end{scope}

\node[bubbletext, fill=green!4!white, inner sep=0.2cm, align=left, text width=7.6cm,
      below=0.5cm of prompt, anchor=north] (tests) {%
      \vspace{-0.5em}
  \begin{minipage}{\linewidth}
  \adjustbox{max width=1\linewidth}{
    \begin{lstlisting}[numbers=none, breaklines=false, keywordstyle=\color{green!50!black}, commentstyle=\color{green!40!black}]
assert rounded_avg(1, 5) == '0b11'
assert rounded_avg(7, 13) == '0b1010'
assert rounded_avg(7, 5) == -1
    \end{lstlisting}
  }
  \end{minipage}};

\node[fill=green!80!black!25, minimum width=8.0cm, minimum height=0.35cm,
      anchor=south, rounded corners=0.15cm] (titlebar2) at ([yshift=0cm]tests.north) {};

\node[text=green!60!black] at (titlebar2.center) {\tiny Test Suite};
\fill[red!70!black]     ([xshift=0.35cm]titlebar2.west) circle (0.06cm);
\fill[yellow!80!orange] ([xshift=0.55cm]titlebar2.west) circle (0.06cm);
\fill[green!60!black]   ([xshift=0.75cm]titlebar2.west) circle (0.06cm);

\draw[rounded corners=0.22cm, draw=gray!30, line width=0.4pt]
  (titlebar.north west) rectangle (prompt.south east);
\draw[rounded corners=0.22cm, draw=green!25!gray, line width=0.4pt]
  (titlebar2.north west) rectangle (tests.south east);


\coordinate (req12) at ($(prompt.east) + (-0.9cm, 0.0cm)$);  
\coordinate (req3)  at ($(prompt.west) + (0.9cm, -1.05cm)$); 

\coordinate (test1) at ($(tests.east) + (-0.8cm, 0.45cm)$);
\coordinate (test2) at ($(tests.east) + (-0.2cm, 0.15cm)$);
\coordinate (test3) at ($(tests.west) + (0.2cm, -0.30cm)$);

\draw[->, thick, teal!70!black, rounded corners=4pt]
  (test1) -- ++(1.0cm,0) |- (req12);
\draw[->, thick, teal!70!black, rounded corners=4pt]
  (test2) -- ++(0.5cm,0) |- (req12);
\draw[->, thick, BurntOrange, rounded corners=4pt]
  (test3) -- ++(-0.6cm,0) |- (req3);

\end{tikzpicture}

%% file: sections/background.tex
\section{Background}



\inlineheadingbf{Large language models} Large language models (LLMs) have become increasingly effective at tasks of code~\cite{DBLP:conf/fose-ws/FanGHLSYZ23, DBLP:journals/corr/abs-2108-07732, DBLP:journals/corr/abs-2107-03374} and test generation~\cite{DBLP:journals/tse/WangHCLWW24, DBLP:journals/pacmse/PizzornoB25, DBLP:journals/corr/abs-2601-09695} from natural language requirements. Here, a developer specifies the intended behavior in natural language and the LLM produces code or test cases reflecting that specification. Formally, given a set of natural language requirement $\mathcal{R}$ encoded in a prompt $\rho$, an LLM generates code by sampling from a probability distribution:
\begin{equation*}
    \mathbb{P}_{LLM}( T \mid \mathcal{R}) = \prod_{i=1}^{n} \mathbb{P}_{LLM}(t_i \mid T_{<i}, \mathcal{R}),
\end{equation*}
where $T = t_1, \dots, t_n$ is an autoregressively generated sequence of tokens and $T_{<i}$ is the prefix at step $i$. The success of LLMs in the coding domain stems from $\mathbb{P}_{LLM}( T \mid \mathcal{R})$ assigning a higher probability to sequence $T$ that correctly implement $\mathcal{R}$. 


\inlineheadingbf{Entropy and Surprise} LLMs are {\em surprised} by (assign a lower probability to) code that does not correctly implement $\mathcal{R}$~\cite{DBLP:conf/icse/RayHGTBD16} or is unrelated to $\mathcal{R}$ altogether~\cite{DBLP:journals/tse/TaherkhaniSTMGH26}. This notion of surprise is formally captured by the cross-entropy of code sequence $T$ given a requirement $\mathcal{R}$.
\begin{equation*}
    H_{LLM}( T \mid \mathcal{R}) = - \frac{1}{n}\sum^n_{i = 1} \log \mathbb{P}_{LLM}(t_i \mid T_{<i}, \mathcal{R})
\end{equation*} 
A higher value of $H_{LLM}( T \mid \mathcal{R})$ indicates that the LLM assigns low probability to the sequence $T$ given $\mathcal{R}$, i.e. the model is more surprised by $T$.

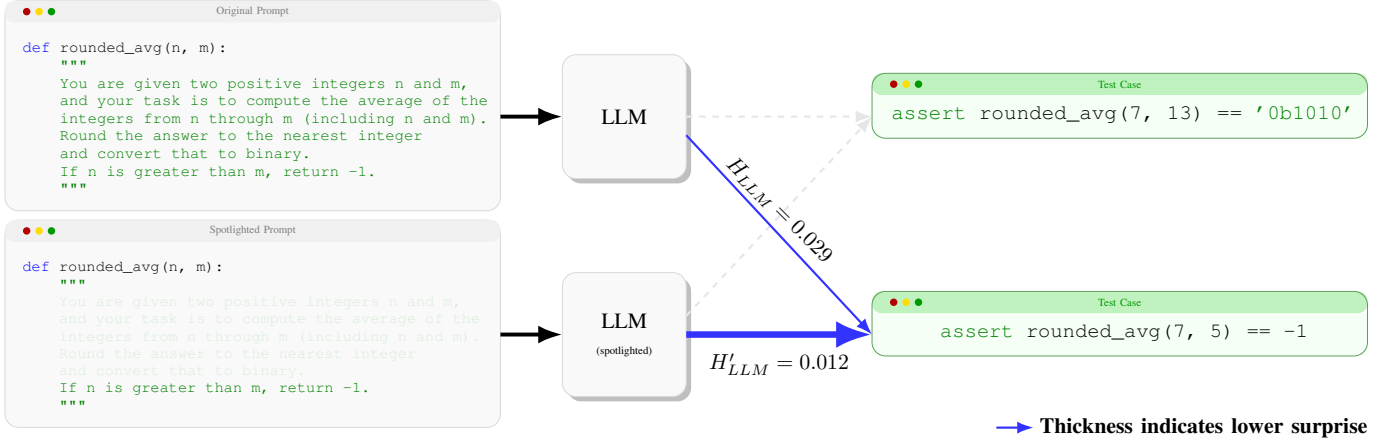
\begin{figure*}[t]
\vspace{1.0em}
\centering
\resizebox{\linewidth}{!}{\input{figures/overview}}
\caption{\textit{Computation of Prompt Coverage Adequacy.} When prompted with a set of requirements, test cases that validate different aspects of the prompt have nearly equal probability, i.e., the LLM is nearly equally surprised by both test cases. By steering the attention of the LLM (spotlighting) towards specific requirement, the LLM becomes more likely to produce (less surprised by) a test that validates that requirement, indicating by the thickness of the blue arrows. We say that a test covers a requirement if after spotlighting the requirement the LLM becomes significantly less surprised by the test. }
\label{fig:overview}
\vspace{1.0em}
\end{figure*}

\inlineheadingbf{Attention} A common architecture of LLMs is the transformer~\cite{DBLP:conf/nips/VaswaniSPUJGKP17}. A transformer consists of $L$ layers, each containing an attention block that captures token relationships, followed by a feed-forward network. Input tokens $t_1, \dots, t_n$ are represented by embeddings $\mathbf{x}_1, \dots, \mathbf{x}_n\in \mathbb{R}^{d}$, where $d$ is the embedding dimension. For each token $t_i$, attention weights over tokens are computed as:
\begin{equation*}
    \alpha_{ij} = \frac{\exp(s_{ij})}{\sum^n_{k = 1}\exp(s_{ik})}, 
\end{equation*}
where $s_{ij} = (\mathbf{x}_iW_{Q})(\mathbf{x}_jW_K)^\top / \sqrt{d}$ is the attention score between $t_i$ and $t_j$, quantifying how much $t_j$ influences the representation of $t_i$. Raw attention weights are naturally interpretable, e.g., if the model is surprised by $t_i$ and the attention score $s_{ij}$ is high, then $t_j$ is likely a contributing factor. However, in a transformer, attention is computed over multiple attention heads and aggregated over $L$ layers, making a direct interpretation of individual scores unreliable~\cite{DBLP:conf/naacl/JainW19}.

\inlineheadingbf{Spotlighting} Spotlighting~\cite{DBLP:conf/eacl/VenkateswaranC26} steers the attention of an LLM towards a designated span of the input. Given a span $\mathcal{S}$ covering the instruction to focus on, the relative attention allocated to the span $\mathcal{S}$ at position $i$ is
$\psi_i = \frac{\sum_{j\in \mathcal{S}} \alpha_{ij}}{ \sum^n_{k = 1} \alpha_{ik}}$.
Let $\psi^*$ be a target attention weight. Whenever $\psi_i < \psi^*$, spotlighting introduces an additive bias $b_{ij}$ to yield a corrected attention weight $\alpha'_{ij}$:
\begin{equation*}
    \alpha'_{ij} = \frac{\exp(s_{ij} + b_{ij})}{\sum^n_{k = 1}\exp(s_{ik} + b_{ik})} \text{ with } b_{ij} = \begin{cases} 
      \log(\frac{\psi^*}{\psi_i}) & j \in \mathcal{S} \\
      0 & \text{other.} 
   \end{cases}
\end{equation*}
By construction, this bias amplifies attention towards $\mathcal{S}$, increasing the influence of the designated instruction on the model's output.


%% file: figures/overview.tex
\begin{tikzpicture}[
    x=1cm,y=1cm,
    every node/.style={inner sep=0pt, outer sep=0pt},
    title/.style={font=\bfseries\fontsize{18}{20}\selectfont},
    sectiontitle/.style={font=\bfseries\fontsize{16}{18}\selectfont},
    bubbletext/.style={font=\ttfamily\fontsize{15}{17}\selectfont, align=center},
    iconlabel/.style={font=\fontsize{22}{24}\selectfont},
]


\node[bubbletext, inner sep=0.2cm, align=left, text width=7.6cm] (prompt) at (0,-1.1) {%
  \adjustbox{max width=\linewidth}{
    \begin{lstlisting}[numbers=none, breaklines=false, escapechar=!]
def rounded_avg(n, m): 
    """
    You are given two positive integers n and m, 
    and your task is to compute the average of the 
    integers from n through m (including n and m).  
    Round the answer to the nearest integer 
    and convert that to binary. 
    If n is greater than m, return -1.
    """
    \end{lstlisting}
}};

\node[fill=gray!12, minimum width=8.0cm, minimum height=0.35cm,
      anchor=south, rounded corners=0.15cm] (titlebar) at ([yshift=0cm]prompt.north) {};

\node[text=gray] at (titlebar.center) {\tiny Original Prompt};

\fill[red!70!black]     ([xshift=0.35cm]titlebar.west) circle (0.06cm);
\fill[yellow!80!orange] ([xshift=0.55cm]titlebar.west) circle (0.06cm);
\fill[green!60!black]   ([xshift=0.75cm]titlebar.west) circle (0.06cm);

\begin{scope}[on background layer]
\draw[fill=gray!5, rounded corners=0.22cm, draw=gray!30, line width=0.4pt]
  (titlebar.north west) rectangle (prompt.south east);
\end{scope}


\node [right=1cm of prompt, fill=gray!5, draw=gray!30, text width=2cm, minimum height=2cm, align=center, rounded corners=0.22cm, drop shadow] (llm1) {LLM};

\draw[-Latex, line width=1.8] (prompt) edge (llm1);


\node[bubbletext, fill=green!4!white, inner sep=0.2cm, align=left, text width=7.6cm,
      right=3cm of llm1] (test1) {%
      \vspace{-0.8em}
  \begin{minipage}{\linewidth}
  \adjustbox{max width=1\linewidth}{
    \begin{lstlisting}[numbers=none, breaklines=false, keywordstyle=\color{green!50!black}, commentstyle=\color{green!40!black}]
assert rounded_avg(7, 13) == '0b1010'
    \end{lstlisting}
  }
  \end{minipage}};

\node[fill=green!80!black!25, minimum width=8.0cm, minimum height=0.35cm,
      anchor=south, rounded corners=0.15cm] (titlebar3) at ([yshift=0cm]test1.north) {};

\node[text=green!60!black] at (titlebar3.center) {\tiny Test Case};
\fill[red!70!black]     ([xshift=0.35cm]titlebar3.west) circle (0.06cm);
\fill[yellow!80!orange] ([xshift=0.55cm]titlebar3.west) circle (0.06cm);
\fill[green!60!black]   ([xshift=0.75cm]titlebar3.west) circle (0.06cm);

\draw[rounded corners=0.22cm, draw=green!25!gray, line width=0.4pt]
  (titlebar3.north west) rectangle (test1.south east);

\node[bubbletext, inner sep=0.2cm, align=left, text width=7.6cm, below=0.5cm of prompt, anchor=north] (prompt2) {%
  \adjustbox{max width=0.75\linewidth}{
    \begin{lstlisting}[numbers=none, breaklines=false, escapechar=!]
def rounded_avg(n, m): 
    """
    !\color{green!50!black!15}You! !\color{green!50!black!15}are! !\color{green!50!black!15}given! !\color{green!50!black!15}two! !\color{green!50!black!15}positive! !\color{green!50!black!15}integers! !\color{green!50!black!15}n! !\color{green!50!black!15}and! !\color{green!50!black!15}m,! 
    !\color{green!50!black!15}and! !\color{green!50!black!15}your! !\color{green!50!black!15}task! !\color{green!50!black!15}is! !\color{green!50!black!15}to! !\color{green!50!black!15}compute! !\color{green!50!black!15}the! !\color{green!50!black!15}average! !\color{green!50!black!15}of! !\color{green!50!black!15}the! 
    !\color{green!50!black!15}integers! !\color{green!50!black!15}from! !\color{green!50!black!15}n! !\color{green!50!black!15}through! !\color{green!50!black!15}m! !\color{green!50!black!15}(including! !\color{green!50!black!15}n! !\color{green!50!black!15}and! !\color{green!50!black!15}m).!  
    !\color{green!50!black!15}Round! !\color{green!50!black!15}the! !\color{green!50!black!15}answer! !\color{green!50!black!15}to! !\color{green!50!black!15}the! !\color{green!50!black!15}nearest! !\color{green!50!black!15}integer! 
    !\color{green!50!black!15}and! !\color{green!50!black!15}convert! !\color{green!50!black!15}that! !\color{green!50!black!15}to! !\color{green!50!black!15}binary.! 
    If n is greater than m, return -1.
    """
    \end{lstlisting}
}};

\node[fill=gray!12, minimum width=8.0cm, minimum height=0.35cm,
      anchor=south, rounded corners=0.15cm] (titlebar2) at ([yshift=0cm]prompt2.north) {};

\node[text=gray] at (titlebar2.center) {\tiny Spotlighted Prompt};

\fill[red!70!black]     ([xshift=0.35cm]titlebar2.west) circle (0.06cm);
\fill[yellow!80!orange] ([xshift=0.55cm]titlebar2.west) circle (0.06cm);
\fill[green!60!black]   ([xshift=0.75cm]titlebar2.west) circle (0.06cm);

\begin{scope}[on background layer]
\draw[fill=gray!5, rounded corners=0.22cm, draw=gray!30, line width=0.4pt]
  (titlebar2.north west) rectangle (prompt2.south east);
\end{scope}


\node [right=1cm of prompt2, fill=gray!5, draw=gray!30, text width=2cm, minimum height=2cm, align=center, rounded corners=0.22cm, drop shadow] (llm2) {LLM \\ {\tiny (spotlighted)}};

\draw[-Latex, line width=1.8] (prompt2) edge (llm2);


\node[bubbletext, fill=green!4!white, inner sep=0.2cm, align=left, text width=7.6cm,
      right=3cm of llm2] (test2) {%
      \vspace{-0.8em}
  \begin{minipage}{\linewidth}
  \adjustbox{max width=0.9\linewidth}{
    \begin{lstlisting}[numbers=none, breaklines=false, keywordstyle=\color{green!50!black}, commentstyle=\color{green!40!black}]
    assert rounded_avg(7, 5) == -1
    \end{lstlisting}
  }
  \end{minipage}};

\node[fill=green!80!black!25, minimum width=8.0cm, minimum height=0.35cm,
      anchor=south, rounded corners=0.15cm] (titlebar4) at ([yshift=0cm]test2.north) {};

\node[text=green!60!black] at (titlebar4.center) {\tiny Test Case};
\fill[red!70!black]     ([xshift=0.35cm]titlebar4.west) circle (0.06cm);
\fill[yellow!80!orange] ([xshift=0.55cm]titlebar4.west) circle (0.06cm);
\fill[green!60!black]   ([xshift=0.75cm]titlebar4.west) circle (0.06cm);

\draw[rounded corners=0.22cm, draw=green!25!gray, line width=0.4pt]
  (titlebar4.north west) rectangle (test2.south east);


\draw[-Latex, line width=1, draw=gray!20, dashed] (llm1) -- (test1);

\draw[-Latex, line width=1, draw=gray!20, dashed] ($(llm2.east) + (0.0, 0.3)$) -- (test1.west);

\draw[-Latex, line width=1, draw=blue!80] ($(llm1.east) + (0.0, -0.3)$) -- node [above=-0.5cm] {\rotatebox[]{-45}{$H_{LLM} = 0.029$}} (test2.west);

\draw[-Latex, line width=3, draw=blue!80] (llm2) -- node [below=0.3cm] {$H'_{LLM} = 0.012$} (test2);


\node [below left=1cm and -8cm of test2] (legend) {\qquad \textbf{Thickness indicates lower surprise}};

\draw[-Latex, line width=1, draw=blue!80] (legend.west) -- ($(legend.west) + (0.6, 0.0)$);

\end{tikzpicture}

%% file: sections/method.tex
\section{Prompt Coverage Adequacy}

We introduce \emph{prompt coverage} as a test adequacy criterion for evaluating a test suite $\mathcal{T}$ with respect to a set of natural language requirements $\mathcal{R} = \{R_1, R_2, \dots, R_n\}$ encoded in a prompt specification. Our coverage criterion rests on two ideas:
\begin{enumerate}
    \item \emph{An LLM should be surprised by a test $T$ that does not meet any requirement in $\mathcal{R}$.} Entropy serves as an alignment metric~\cite{DBLP:journals/tse/TaherkhaniSTMGH26} where higher entropy indicates lower alignment between test case $T$ and the natural language requirement $\mathcal{R}$.
    \item \emph{Focusing the attention of the LLM on an aligned requirement $R_i \in \mathcal{R}$ should reduce surprisal.} Spotlighting $R_i$ amplifies its influence on the output entropy: if $R_i$ is validated by $T$, surprisal should decrease relative to the baseline.
\end{enumerate}
We see the latter as evidence that $T$ covers the behavior specified by $R_i$. An overview is shown in \Cref{fig:overview}.

\subsection{Prompt Coverage Criterion} 
Given a prompt $\rho$ encoding a set of natural language requirements $\mathcal{R}$ and a test suite $\mathcal{T}$, our goal is to identify which requirements $R_i \in \mathcal{R}$ are validated ({\em covered}) by tests in $\mathcal{T}$. 

\inlineheadingbf{Coverage criterion} $R_i$ as covered by $T$ iff:
\begin{equation}\label{eq:covering_crit}
    H_{LLM}(T\mid \dots R^+_i\dots) - H_{LLM}(T\mid \mathcal{R}) < \tau,
\end{equation}
where $\tau \leq 0$ is an acceptance threshold. $H_{LLM}(T\mid \mathcal{R})$ denotes the baseline per-token entropy of $T$ given the full prompt, and $H_{LLM}(T\mid  \dots R^+_i \dots)$ the entropy after spotlighting $R_i$. The key insight is that spotlighting $R_i$ forces the LLM to attend more to $R_i$ when evaluating $T$. If $R_i$ and $T$ are semantically aligned, this additional focus provides useful context and the model becomes less surprised by $T$. Conversely, if $R_i$ is unrelated to $T$, the forced attention distracts the model from more relevant requirements, and surprisal increases. This can be interpreted as a soft form of \emph{counterfactual reasoning}~\cite{DBLP:journals/csur/VermaBHHDS24}: had the LLM attended more to $R_i$, it would have assigned higher or equal probability to $T$.

\inlineheadingbf{Prompt coverage} The {\em prompt coverage set} of $\mathcal{T}$ is the subset of requirements covered by at least one test in $\mathcal{T}$:
\begin{equation*}
    \text{Cov}_{\mathcal{R}}(\mathcal{T}) = \{R_i \in \mathcal{R}\mid \exists T\in \mathcal{T}: R_i \text{ is covered by } T \}
\end{equation*}
Requirements $R_i \not\in Cov_{\mathcal{R}}(\mathcal{T})$ are {\em uncovered}, signaling gaps that require additional testing. The \emph{prompt coverage score} $PC(\mathcal{T}, \mathcal{R}) = |\text{Cov}_{\mathcal{R}}(\mathcal{T})|/|\mathcal{R}|$ measures the proportion of requirements covered by $\mathcal{T}$, with score of 1.0 indicating full coverage, according to the LLM.

\begin{algorithm}[t]
\caption{Prompt coverage sets of a test suite $\mathcal{T}$}
\label{alg:coverage}
\begin{algorithmic}[1]
\Require Prompt $\rho$ (encoding $\mathcal{R}$), language model $LLM$, a test suite $\mathcal{T}$, an acceptance threshold $\tau$
\Ensure Prompt Coverage Set $Cov_\mathcal{R}(\mathcal{T})$

\State $\mathcal{R} \gets \text{segment}(\rho)$ \Comment{Segment $\rho$ into requirements}
\State covered $\gets \emptyset$
\For{$T \in \mathcal{T}$}
    \State $baseline \gets H_{LLM}(T\mid \mathcal{R})$
    \For{$R_i \in \mathcal{R} \setminus covered$}
        \State $surprise_i \gets H_{LLM}(T\mid \dots R^+_i \dots)$
        \If{$surprise_i - baseline < \tau$}
            \State $covered \gets covered \cup \{R_i\}$
        \EndIf
    \EndFor
\EndFor
\Return $covered$
\end{algorithmic}
\end{algorithm}

\subsection{Computing Prompt Coverage Sets}
\Cref{alg:coverage} provides an overview of our algorithm for computing prompt coverage sets. 

\inlineheadingbf{Prompt Segmentation} We start by segmenting the code generation prompt $\rho$ into individual requirements $\mathcal{R} = \{R_1, \dots, R_n\}$ using a rule-based parser that identifies natural language requirement sentences among other prompt elements such as function signatures, example test cases, and type annotations. Each extracted $R_i$ corresponds to an atomic natural language requirement.

\inlineheadingbf{Coverage computation} For each test $T \in \mathcal{T}$, we compute the baseline entropy $H_{LLM}(T\mid \mathcal{R})$ by evaluating the LLM on $T$ conditioned on the full prompt. We then spotlight each uncovered requirement $R_i$ in turn, computing the spotlighted entropy $H_{LLM}(T\mid \dots R^+_i \dots)$. If the spotlighted entropy does not exceed the baseline, $R_i$ is added to the coverage set. Requirements already covered are skipped, as coverage is monotone: once a requirement is covered by some $T \in \mathcal{T}$, no further tests need to be evaluated against it.

\subsection{Implementation}
\inlineheadingbf{Entropy computation} For long test cases, per-token entropy can become unreliable due to length-dependent accumulation of uncertainty. We therefore approximate entropy by reducing it to a single-token prediction: 
\begin{equation*}
    H_{LLM}(T \mid \mathcal{R}) \approx H_{LLM}( \text{\faCheck} \mid \rho(\mathcal{R}, T)),
\end{equation*}
where $\rho(\mathcal{R}, T)$ is a prompt asking the LLM whether $T$ is entailed by $\mathcal{R}$ and \faCheck~ is a single token indicating entailment. A lower probability of \faCheck~ indicates that the LLM finds $T$ less consistent with $\mathcal{R}$, coresponding to a higher surprise.

\inlineheadingbf{Noise in entropy} In practice, we would want to set $\tau \approx 0$ to identify requirements $R_i$ that reduce the model's surprise. However, spotlighting introduces noise in the process by manipulating the attention weights, resulting in some cases in marginal entropy variations that do not reflect genuine changes in requirement coverage. To address this, we tuned $\tau$ on a small set of training examples to find a threshold that gives meaningful coverage reports.


%

%% file: sections/evaluation.tex
\section{Evaluation}

\subsection{Research Questions}
We aim to assess the suitability of prompt coverage as a criterion for evaluating test suite effectiveness and guiding the generation of high-quality test suites. To this end, we investigate the following research questions:
\begin{description}
\item[\textbf{RQ1}] To what extend prompt coverage relates with traditional code coverage criteria?
\item[\textbf{RQ2}] To what extend prompt coverage relates with fault detection?
\item[\textbf{RQ3}] How does prompt coverage-driven test generation compares with code coverage-driven one in terms of fault detection?
\end{description}
\textbf{RQ1} evaluates whether prompt coverage, computed at the prompt level, relates with code coverage computed on implementations derived from the prompt. RQ2 and RQ3 evaluate the relevance of the criterion by examining whether it can help build more effective test suites than code coverage.

\subsection{Benchmarks}

To evaluate our proposed criteria, we selected two widely-used coding benchmarks: HumanEval+ \cite{DBLP:conf/nips/LiuXW023} and LiveCodeBench \cite{DBLP:conf/iclr/JainHGLYZWSSS25}. HumanEval+ is composed of 164 Python programming problems. LiveCodeBench is a benchmark of Coding Contest tasks which are periodically updated to prevent test contamination. We used the latest version of LiveCodeBench (v6) that is composed of 112 tasks collected in 2025. We selected these benchmarks among available coding benchmarks as they have well-formed prompts with multiple requirements and comprehensive test suites that validate these requirements. 

\inlineheadingit{Implementation candidates} To compute {\em code} coverage and fault detection, we augment each benchmark with coding solutions sampled from an LLM. Incorrect solutions (as judged by the benchmark's test suite) are treated as faulty implementations that fail to satisfy one or more requirements, while solutions passing the test suite are considered correct. We employ Qwen2.5-Coder to generate 10 candidate solutions per benchmark task. To balance correct and incorrect solutions, we use the weaker Qwen2.5-Coder (7B) model for HumanEval+ and the stronger Qwen2.5-Coder (32B) for LiveCodeBench. This choice reflects the relative difficulty of the two benchmarks: HumanEval+ is comparatively easy, so a weaker model is more likely to produce a mix of correct and incorrect solutions, whereas LiveCodeBench is harder and requires a stronger model to yield a sufficient number of correct solutions.

\inlineheadingit{Test Selection} As coverage quickly satures among the coding examples, we subsample around 20 -- 30 test cases per benchmark tasks. For LiveCodeBench, we excluded performance tests (with overly large inputs), as we are mainly interested in testing functional requirements.


\subsection{Coverage criteria}

\inlineheadingit{Statement coverage} We consider statement coverage as our main baseline, a well-established criterion widely used to guide modern testing practices. To measure statement coverage, we employ the Python \texttt{coverage} tool, while running tests via \texttt{pytest}. \texttt{coverage} reports which statements (lines of code) were executed at least once during test execution. The statement coverage score (SC) can only be computed on code implementing the requirement, but not the requirement itself. Therefore, we report the statement coverage score as the percentage of lines covered of a particular implementation.

\inlineheadingit{Prompt coverage (ours)} We evaluate prompt coverage with respect to two popular coding models: Qwen2.5-Coder (14B) and Gemma3 (12B). These models have achieved strong coding performance and are small enough to fit on a single consumer GPU, accounting for the overhead due to attention. To compute prompt coverage, we have adapted the implementation of \textit{Spotlight}~\cite{DBLP:conf/eacl/VenkateswaranC26} for our use case, setting the parameter $\psi^* = 0.1$. We excluded closed-source models from our evaluation, as these do not allow the computation of entropy nor can be guided via spotlighting.


\subsection{Statistical Correlations}
For RQ1 and RQ2, we perform a statistical analysis between prompt coverage and code coverage or fault detection respectively. To do so, we fitted a Beta generalized linear mixed model (GLMM) \cite{GLMM} via \textit{glmmTMB} \cite{glmmTMB} with a logit link to assess their relationship. For fault detection, we used a binomial generalized linear mixed model given the binary nature of fault detection. Furthermore, given that coverage is a value in [0, 1] and tend to saturate around 1.0, the Beta family was chosen to respect the [0,1] bound and the logit link to capture saturation at the ceiling. On top of prompt coverage as a fixed effect, we included the test subset size, and their interaction. Adding the subset size was done to control whether we could trivially increase coverage simply by increasing the subset size. The interaction effect was added to control the saturation effect and diminishing return at high subset size: if a sample already tests many examples, improving prompt coverage has less room to further increase code coverage. We further added random intercept per task as our observations are not truly independent: multiple code samples come from the same task. This results in the following equation: $\beta_0 + \beta_1 \times PC + \beta_2 \times subset\_size + \beta_3 \times PC:subset\_size + u_{task}$, with $u_{task} \sim N(0, \tau^2)$ the random intercept per task and $PC$ the Prompt Coverage. We fit the statistical models over runnable solution candidates where a meaningful code coverage and fault detection score can be computed, excluding those that do not run on any test input.

%% file: sections/results.tex
\section{Results}

\subsection{RQ1: Prompt Coverage and Code Coverage}

\begin{table*}[htbp]
\centering
\caption{Generalized Linear Mixed-Effects Model (GLMM) studying the effect of prompt coverage over code coverage. All $p$-values are significant at threshold $p < 0.01$. Est = Estimate, SE = Standard Error, OR = Odds Ratio.}
\label{tab:rq1_glmm_results}
\small 
\begin{tabular}{l ccc ccc c ccc ccc}
\toprule
& \multicolumn{6}{c}{\textbf{HumanEval+}} & & \multicolumn{6}{c}{\textbf{LiveCodeBench}} \\
\cmidrule(lr){2-7} \cmidrule(lr){9-14}
& \multicolumn{3}{c}{Qwen2.5} & \multicolumn{3}{c}{Gemma3} & & \multicolumn{3}{c}{Qwen2.5} & \multicolumn{3}{c}{Gemma3} \\
\cmidrule(lr){2-4} \cmidrule(lr){5-7} \cmidrule(lr){9-11} \cmidrule(lr){12-14}
Predictor & Est & SE & OR & Est & SE & OR & & Est & SE & OR & Est & SE & OR \\
\midrule
(Intercept)      & $\phantom{-}1.29$ & 0.06 & 3.62  & $\phantom{-}1.93$ & 0.05 & 6.88 & & $\phantom{-}1.38$ & $0.04$ & $3.99$ & $\phantom{-}1.63$ & $0.05$ & $5.12$ \\
prompt\_coverage & $\phantom{-}1.84$ & 0.05 & 6.30 & $\phantom{-}1.21$ & 0.04 & 3.35 & & $\phantom{-}0.82$ & 0.04 & 2.26 & $\phantom{-}0.47$ & 0.03 & 1.60 \\
subset\_size     & $\phantom{-}0.12$ & 0.02 & 1.13  & $\phantom{-}0.11$ & 0.01 & 1.12 & & $\phantom{-}0.06$ & 0.01 & 1.06 & $\phantom{-}0.05$ & 0.00 & 1.05 \\
PC:subset\_size  & $-0.08$   & 0.02 & 0.92  & $-0.07$ & 0.01 & 0.93 & & $-0.05$ & 0.01 & 0.95 & $-0.02$ & 0.00 & 0.98 \\
\bottomrule
\end{tabular}
\vspace{0.2em}
\end{table*}

\begin{figure}
    \centering
    \vspace{-0.2em}
    \includegraphics[width=1.0\columnwidth]{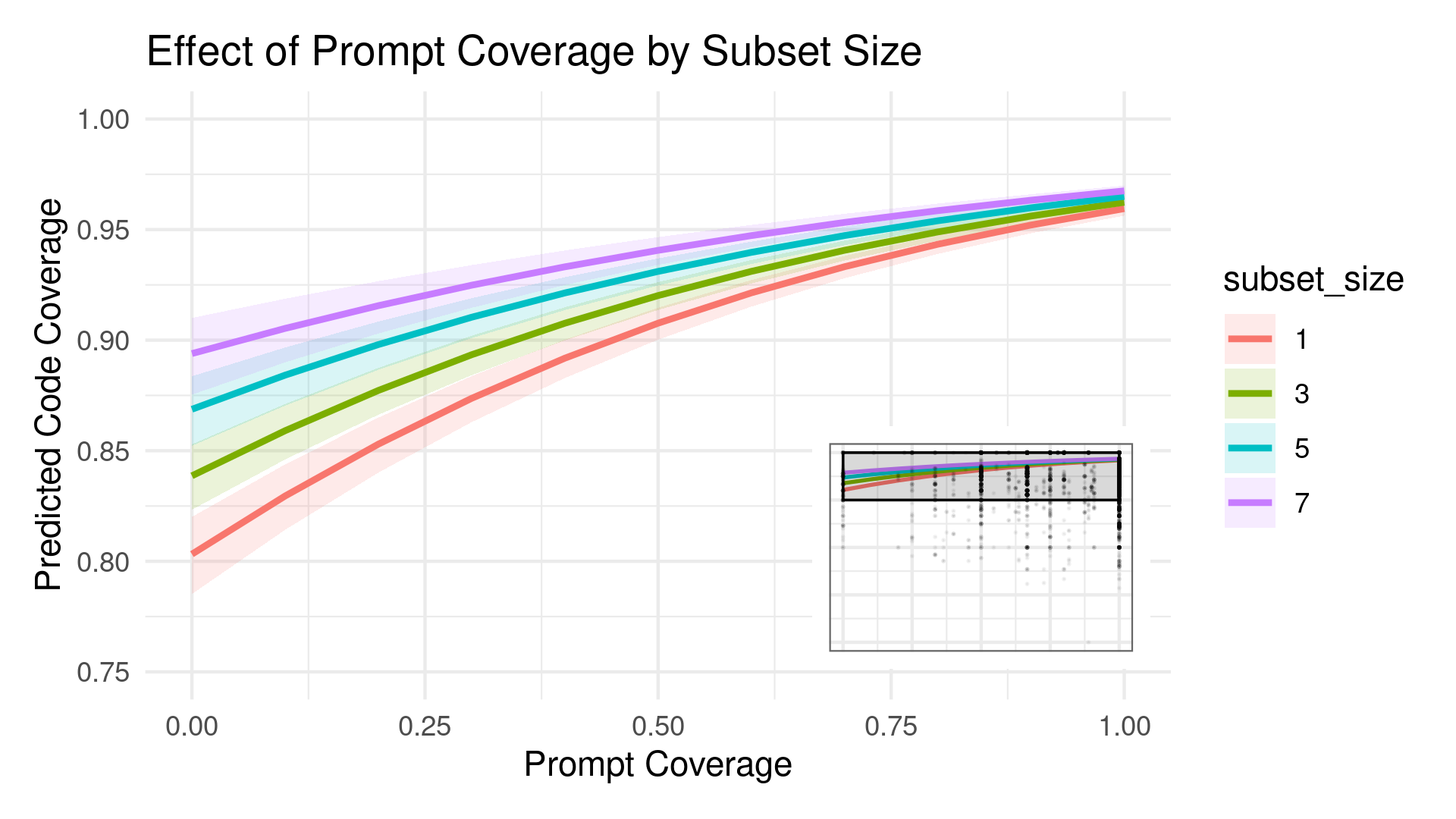}
   \vspace{-0.5em}
    \caption{Prompt Coverage effect over Code Coverage as described by the model controlling for subset size. Coverage is zoomed in as coverage saturate easily. Code Coverage increases with Prompt Coverage, with diminishing returns as subset size increases.}
    \label{fig:rq1_cc_vs_pc}
    \vspace{-0.2em}
\end{figure}

\Cref{tab:rq1_glmm_results} shows the result of the analysis by the GLMM model. Furthermore, \Cref{fig:rq1_cc_vs_pc} illustrates the trend for prompt coverage computed by Qwen-2.5-Coder on HumanEval+. We observe similar trends in the other cases.

\inlineheadingit{HumanEval+} For both Qwen2.5-Coder and Gemma3, prompt coverage (PC) is a strong, significant predictor of code coverage (Qwen2.5: $\beta$=1.84, OR=6.30, p $<$ 0.001; Gemma3: $\beta$=1.21, OR=3.35, p $<$ 0.001). The interaction effect of PC and test subset\_size is significant but show small interaction (Qwen2.5: $\beta$=-0.083, OR=0.92; Gemma3: $\beta$=-0.071, OR=0.93; both p $<$ 0.001) which confirms that the effect of prompt coverage attenuates as the test subset grows, albeit slightly. Test subset size on its own has a positive but modest main effect (OR $\simeq$ 1.12), indicating that larger subsets produce marginally higher coverage regardless of prompt quality. In practice, a +10pp increase in prompt coverage can be achieved by adding roughly 2.5 tests for Qwen2.5-Coder and 1 test for Gemma3 to a test suite that already contain three tests (subset\_size 3). This benefit shrinks to less than +0.6 tests at the largest sizes.


\inlineheadingit{LiveCodeBench} On LiveCodeBench, we observe a similar, but weaker, trend: PC still has a positive and significant effect (Qwen2.5: $\beta = 0.82$, OR = 2.26, p $<$ 0.001; Gemma3: $\beta = 0.47$, OR = 1.60, p $<$ 0.001), subset\_size has a mostly neutral effect (OR $\simeq$ 1.05), and the diminishing effect with subset\_size is also present but weaker (Qwen2.5: $\beta$=-0.05, OR=0.95; Gemma3: $\beta$=-0.03, OR=0.98; both p $<$ 0.001). In practice, in both case, a +10 pp in PC equates to +1.0 extra tests at subset\_size 3. This can be partly explained by the fact that LiveCodeBench is harder (30\% correct samples vs $\simeq$ 80\% for HumanEval+): Coding contest tasks are often embedded within a narrative, which makes it significantly harder to align tests with the underlying natural language requirements.
Nonetheless, we still observe the same link between prompt and code coverage, a link which cannot be explained mainly by the subset size.

\begin{findingbox}[frametitle={Finding RQ1}]
Prompt coverage and code coverage are linked: the more prompt coverage increases, the more code coverage does. This behavior is not dependent on the subset size only, and at equivalent subset size, a higher prompt coverage will lead to more of the implementation being covered, despite the fact that prompt coverage is independent of the implementation.
\end{findingbox}

\subsection{RQ2: Prompt Coverage and Fault Detection}

\begin{table*}[htbp]
\centering
\caption{Generalized Linear Mixed-Effects Model (GLMM) studying the effect of prompt coverage over fault detection. All $p$-values are significant at threshold $p < 0.01$ unless specified otherwise (${\dagger} = p < 0.05$, ${\times} =$ not significant). Est = Estimate, SE = Standard Error, OR = Odds Ratio.}
\label{tab:rq2_glmm_results}
\small 
\begin{tabular}{l ccc ccc c ccc ccc}
\toprule
& \multicolumn{6}{c}{\textbf{HumanEval+}} & & \multicolumn{6}{c}{\textbf{LiveCodeBench}} \\
\cmidrule(lr){2-7} \cmidrule(lr){9-14}
& \multicolumn{3}{c}{Qwen2.5} & \multicolumn{3}{c}{Gemma3} & & \multicolumn{3}{c}{Qwen2.5} & \multicolumn{3}{c}{Gemma3} \\
\cmidrule(lr){2-4} \cmidrule(lr){5-7} \cmidrule(lr){9-11} \cmidrule(lr){12-14}
Predictor & Est & SE & OR & Est & SE & OR & & Est & SE & OR & Est & SE & OR \\
\midrule
(Intercept)      & $-2.20$ & 0.51 & 0.11  & $-1.67$ & 0.56 & 0.19 & & $\phantom{-}0.60^{\times}$ & $0.39$ & $1.82$ & $\phantom{-}0.65^{\dagger}$ & $0.29$ & $1.91$ \\
prompt\_coverage & $\phantom{-}2.54$ & 0.37 & 12.73 & $\phantom{-}1.18$ & 0.21 & 3.27 & & $\phantom{-}1.40$ & 0.15 & 4.04 & $\phantom{-}1.47$ & 0.12 & 4.34 \\
subset\_size     & $\phantom{-}0.54$ & 0.14 & 1.72  & $\phantom{-}0.48$ & 0.05 & 1.62 & & $\phantom{-}0.46$ & 0.05 & 1.59 & $\phantom{-}0.40$ & 0.03 & 1.49 \\
PC:subset\_size  & $-0.29^{\dagger}$   & 0.14 & 0.75  & $-0.19$ & 0.06 & 0.83 & & $-0.18$ & 0.05 & 0.84 & $-0.14$ & 0.03 & 0.87 \\
\bottomrule
\end{tabular}
\end{table*}

\begin{figure}
    \centering
    \includegraphics[width=0.97\columnwidth]{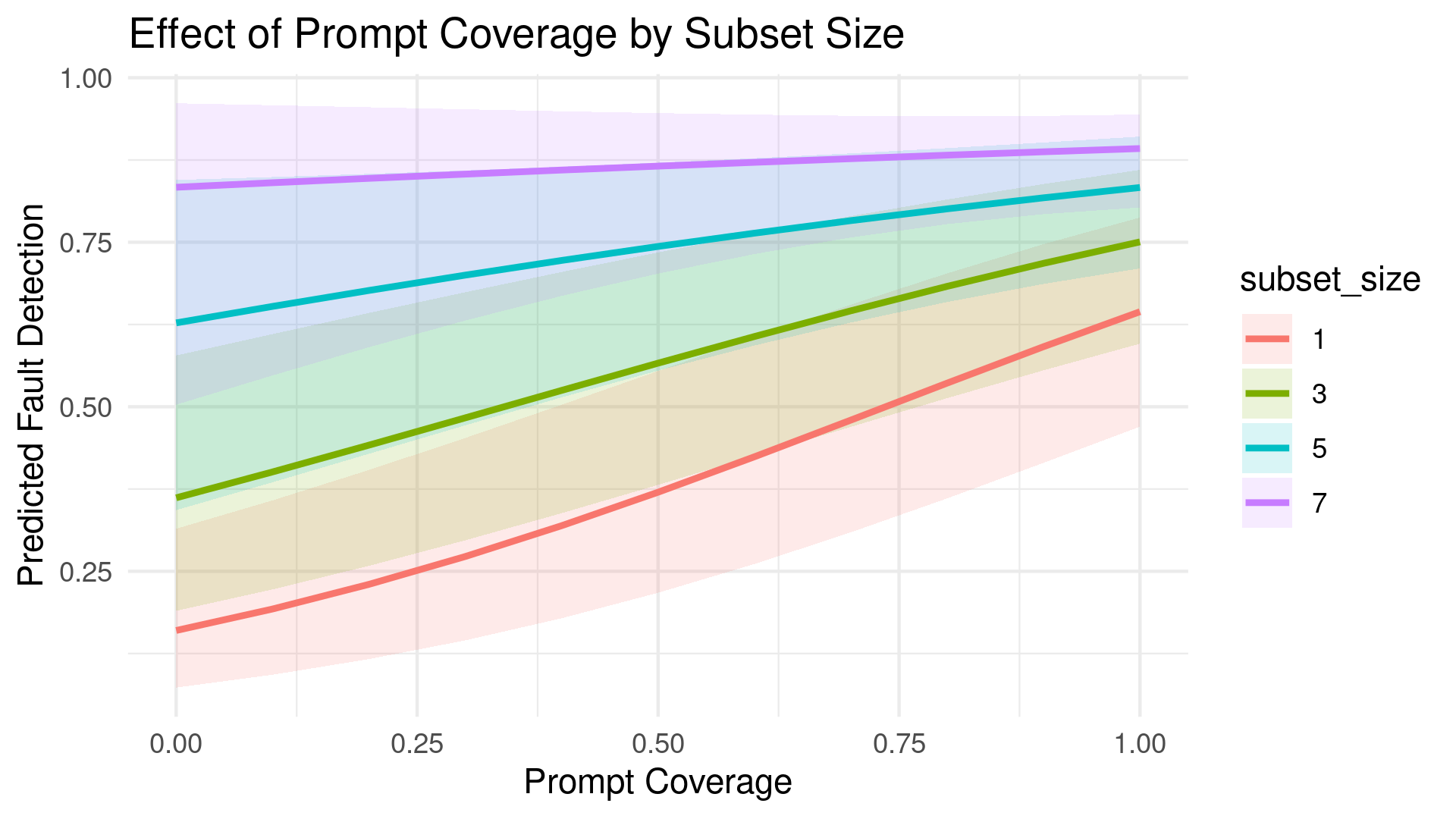}
    \vspace{-0.5em}
    \caption{Prompt Coverage effect over Fault detection as described by the model controlling for subset size. Fault detection increases with Prompt Coverage, with diminishing returns as subset size increases.}
    \label{fig:rq2_fd_vs_pc}
    \vspace{-0.3em}
\end{figure}

\Cref{tab:rq2_glmm_results} shows the parameters of the GLMM model fitted for fault detection. \Cref{fig:rq2_fd_vs_pc} illustrates the observed correlation between prompt coverage computed with Qwen2.5-Coder and fault detection on HumanEval+. Similar trends can be observed in the other cases.

\inlineheadingit{HumanEval+} We show that prompt coverage significantly increases fault detection within a HumanEval task (Qwen2.5: $\beta = 2.54$, OR = 12.73, p $<$ 0.001; Gemma3: $\beta = 1.18$, OR = 3.27, p $<$ 0.001). While test subset size has a positive effect (OR $\approx$ 1.62–1.72), it is still lower than the effect of PC. The interaction effect of PC and subset\_size (PC:subset\_size) has negative interaction and is significant in both models (Qwen2.5: $\beta = -0.29$, OR = 0.75, p = 0.035; Gemma3: $\beta = -0.19$, OR = 0.83, p $<$ 0.001), indicating diminishing returns similarly to RQ1. At small subset\_size (3), a continously increase of prompt coverage from 0.0 to 1.0 increases detection probability by +39pp (Qwen2.5) and +15pp (Gemma3), but the effect shrinks substantially as more tests are added. 

\inlineheadingit{LiveCodeBench} On LiveCodeBench, prompt coverage significantly increases fault detection, with comparable effects across both models (Qwen2.5: $\beta = 1.40$, OR = 4.04, p $<$ 0.001; Gemma3: $\beta = 1.47$, OR = 4.34, p $<$ 0.001). Subset size also has a positive but lower effect (OR $\approx$ 1.49–1.59): as such, similarly to RQ1, the main effect comes from prompt coverage and not the subset size. The interaction of PC and subset\_size is negative and significant (Qwen2.5: $\beta = -0.18$, OR = 0.84, p $<$ 0.001; Gemma3: $\beta = -0.14$, OR = 0.87, p $<$ 0.001), confirming diminishing returns, albeit at a smaller rate than on HumanEval+. At small subset\_size (3), increasing PC from 0.0 to 1.0 raises detection probability by +7pp (Qwen2.5) and +9pp (Gemma3), but the effect vanishes at a higher subset size due to the initial high baseline detection rate. 

\begin{findingbox}[frametitle={Finding RQ2}]
Increasing prompt coverage leads to a higher chance of detecting faults. This effect is primarily attributed to the objectives of prompt coverage and not an artifact of the test subset size. Our results demonstrate that even when controlling for subset size, higher prompt coverage leads to higher fault detection. 
\end{findingbox}

\begin{figure}[!tb]
    \centering
    \begin{subfigure}[b]{0.45\columnwidth}
        \centering
        \includegraphics[width=\columnwidth]{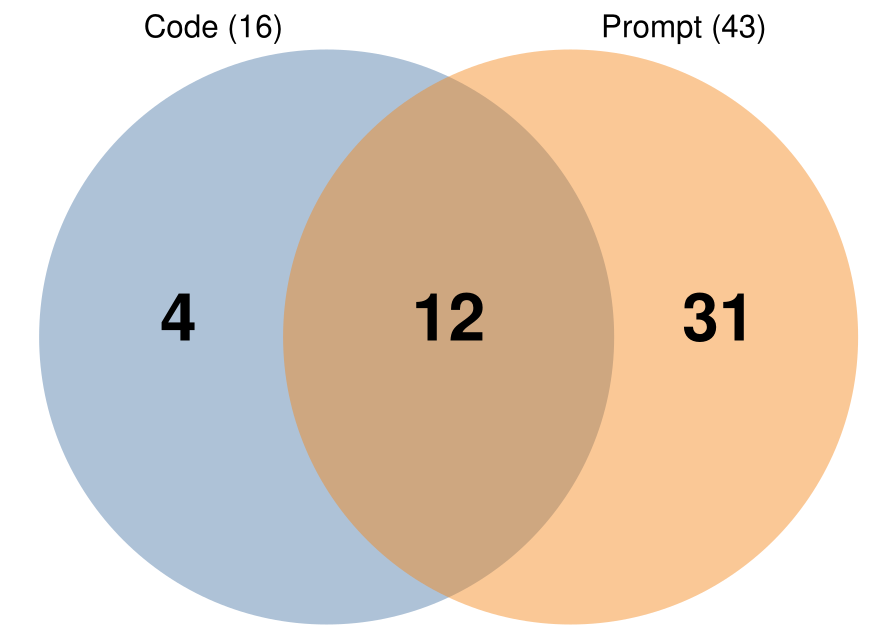} 
    \end{subfigure}
    \hfill
    \begin{subfigure}[b]{0.45\columnwidth}
        \centering
        \includegraphics[width=\columnwidth]{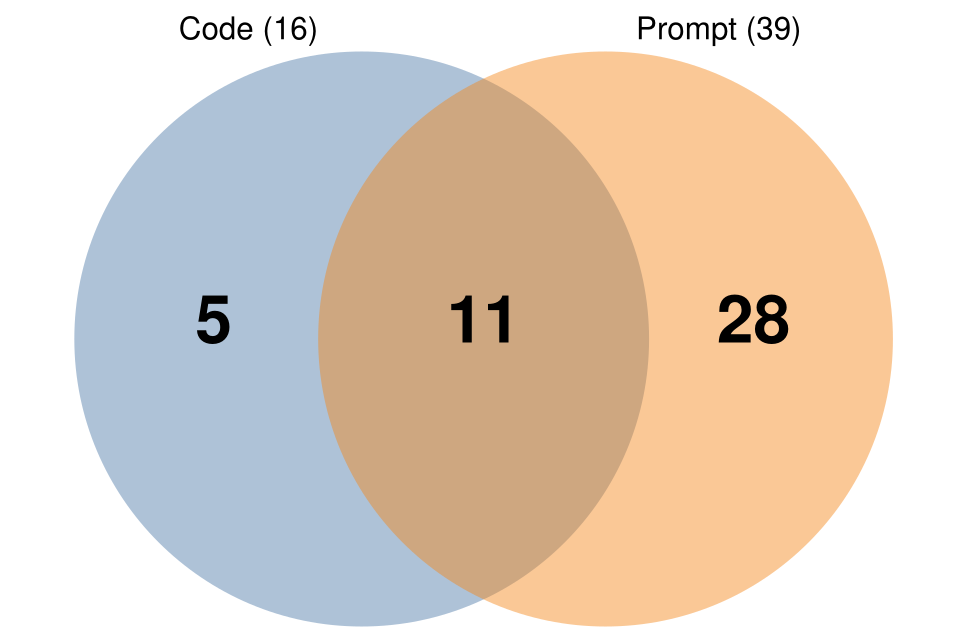} 
    \end{subfigure}
    
    \caption{Venn diagram of faults detected per each method: code guided (blue) or prompt guided (orange). Data for HumanEval+ (left) or LiveCodeBench (right).}
    \label{fig:rq3}
\end{figure}

\subsection{RQ3: Coverage-Driven Testing}
We now evaluate whether our findings have practical implications for testing LLM-generated code. To this end, we evaluate the fault detection capabilities of coverage-driven testing guided by our prompt coverage criterion. 

\inlineheadingit{Testing Protocol} We select benchmark tasks with incorrect solutions where the fault is not covered by a randomly selected subset of tests. We then use GPT-4.1-nano to augment the test suite, either guided by prompt coverage or code coverage. We encode coverage criteria into the prompt: For code coverage, the LLM is provided with the incorrect implementation where uncovered lines are marked. For prompt coverage, we annotate the prompt itself with requirements not covered by the current test suite. A fault is detected if the generated tests pass on correct solutions but fail on the given faulty implementation. In total, we were able to collect faulty implementation for 88 HumanEval tasks and 68 LiveCodeBench tasks.


\inlineheadingit{Results} Our fault detection results are shown in \Cref{tab:test_gen}.
Augmenting the test suite with coverage-guided tests improves fault detection considerably: Guided by code coverage, the LLM-generated tests only detect 18.2\% of HumanEval faults and 23.5\% of LiveCodeBench faults despite triggering roughly $\times$ 3 the number of faults. By using our prompt coverage criterion, the LLM-generated tests detects 48.9\% and 57.4\% of faults on HumanEval and LiveCodeBench respectively and potentially trigger up to 92.0\% and 76.5\% of the faults on HumanEval+ and LiveCodeBench. This is an increase of +30.7pp on HumanEval and +33.9pp on LiveCodeBench in fault detection. The passing rate, i.e. the percentage of tests that pass on both faulty and correct implementations, is comparable between both test augmentation methods.

These findings highlight a fundamental difference between code coverage and prompt coverage. When LLM is guided by code coverage, it targets the behavior implemented in the code. In contrast, when guided by prompt coverage, it tests what is defined in the prompt, making it possible to detect behavior that deviates from the stated requirements.

Still, \Cref{fig:rq3} shows that there are complementary advantages of coverage-driven testing guided by code and prompt coverage. While prompt coverage more effectively guides the LLM, yielding 31 (HumanEval+) and 28 (LiveCodeBench) faults uniquely found by prompt coverage-guided testing, code coverage-guidance helps to write tests that uncover 4 and 5 unique faults.  





\begin{findingbox}[frametitle={Finding RQ3}]
Coverage-driven testing guided by prompt coverage uncovers 30+\% more faults than code coverage-guided testing. 
Our results highlight the advantage of prompt coverage over traditional code coverage criteria, capturing aspects of the program behavior that are missed by code coverage and that can guide LLMs effectively in generating fault-detecting tests.
\end{findingbox}

\begin{table*}[h]
    \centering
    \vspace{0.5em}
    \caption{Results of coverage-driven test augmentation guided by prompt coverage or code coverage. For both datasets and methods we report: (1) proportion of tests that pass on both faulty implementation AND the correct implementations, (2) percentage (and raw number) of faulty implementations for which at least one test leads to a behavior that is different between the two implementations, i.e., triggers the fault, 3) percentage (and raw number) of faulty implementations for which at least one test FAIL on the faulty code but PASS on the reference implementation, fault is triggered.}
      \vspace{0.5em}
    \renewcommand{\arraystretch}{1.2}
    \resizebox{\textwidth}{!}{%
    \begin{tabular}{l rrr c rrr}
    \toprule
         & \multicolumn{3}{c}{\bfseries Code Coverage} &&  \multicolumn{3}{c}{\bfseries Prompt Coverage (ours)}\\
         \cmidrule{2-4} \cmidrule{6 - 8}
         {\bfseries Benchmark} & \% valid test & \% / \# faults triggered & \% / \# faults detected && \% valid test & \% / \# faults triggered & \% / \# faults detected\\
         \midrule
         HumanEval+ & 75.1\% & 61.4\% / 54 & 18.2\% / 16 && 77.8\% & 92.0\% / 81 & 48.9\% / 43 \\
         LiveCodeBench & 63.5\% & 51.5\% / 35 & 23.5\% / 16 && 70.0\% & 76.5\% / 52 & 57.4\% / 39 \\
         \bottomrule
    \end{tabular}}
          \vspace{0.5em}
    \label{tab:test_gen}
\end{table*}

%% file: sections/threats.tex
\section{Threats to Validity}\label{sec:threats}

 While prompt coverage is subject to the limitations inherent in any new adequacy criterion, our results provide encouraging evidence that prompt coverage captures meaningful information about test effectiveness. Future work involving additional benchmarks, programming languages, model families, and alternative coverage formulations will help further assess its validity and applicability in LLM-driven software development. In the following, we discuss validity threats that may affect our findings.

\paragraph{Construct Validity} A threat to construct validity arise from the definition of the prompt coverage criterion. As it is based on changes in model entropy induced by selectively increasing attention toward candidate natural language test requirements, it is possible that entropy variation only partially capture the extent to which a test suite explores a prompt's semantic space. Other attention or representation-based measures could provide alternative interpretations of prompt coverage. 

Additionally, our approach assumes that the attention mechanisms of the underlying models provide meaningful signals regarding the relevance of prompt requirements. Although attention has been widely used to analyze model behavior, it may not reflect the reasoning process of the model. Consequently, prompt coverage should be viewed as an operationalization of the ideal coverage metric rather than a complete representation of specification coverage. Another threat concerns the use of fault detection as a proxy for test effectiveness. While fault detection is a standard evaluation criterion in software testing research, undetected faults may still exist, and improvements in fault detection may not reflect all aspects of test utility. 

\paragraph{Internal Validity} Threats to internal validity concern factors that may influence the observed relationship between prompt coverage, code coverage, and fault detection. One potential threat is the configuration of the test generation process. Variations in prompts, sampling parameters, temperature settings, or model specific behaviors may affect both the generated tests and the resulting coverage measurements. Similarly, the selected implementations may contain characteristics that favor certain testing strategies. Although our experiments compare prompt coverage-guided and code coverage-guided generation under the same experimental conditions, uncontrolled factors in the generation process could contribute to the observed differences.

Another threat regards the computed associations between prompt coverage and code coverage that may not necessarily imply causation. It is possible that both metrics are influenced by underlying factors related to prompt and code complexity or test diversity. Additionally the size of our experimental dataset(s) may also threaten our conclussions. Although HumanEval and LiveCodeBench provide a substantial number of programming tasks, larger scale evaluations may reveal additional patterns, or encounter different distributions and include different edge cases that we might miss. 

Moreover, the reported improvement of up to 30\% in fault detection represents the behavior observed on our evaluation benchmarks and may vary across other datasets or experimental settings. Despite these limitations, the consistency of our findings across two benchmarks and two different LLMs increases confidence that the observed effects are not model or dataset-specific artifacts. Nevertheless, further studies on larger and more diverse workloads will be necessary to fully establish the robustness and general applicability of prompt coverage.

\paragraph{External Validity} The generalizability of our results may be limited by the choice of datasets and language models. We evaluated our prompt coverage criterion on HumanEval and LiveCodeBench, which are widely used benchmarks for code generation. However, these benchmarks primarily consist of programming tasks with relatively concise natural language specifications. The behavior of prompt coverage may differ for industrial software systems, long requirements documents, or complex projects. 

Similarly, our study considers two contemporary open source models, Qwen2.5 and Gemma3. Although these models represent different architectures and training regimes, the results may not generalize to all proprietary or future LLMs. Since prompt coverage relies on model attention and entropy measurements, effectiveness may vary across model families. We anyway had to use  open source models to be able to have access to the model attention and entropy. 

In addition, our experiments focus on Python code generation tasks. Whether the same relationships between prompt coverage, code coverage, and fault detection hold for other programming languages or software engineering tasks remains an open question.

%% file: sections/related.tex
\section{Related Work}
\subsection{Large language models}
Large language models are increasingly used by engineers to perform various software engineering tasks, such as code generation~\cite{DBLP:conf/fose-ws/FanGHLSYZ23, DBLP:journals/corr/abs-2108-07732, DBLP:journals/corr/abs-2107-03374}, automated testing~\cite{DBLP:journals/tse/WangHCLWW24, DBLP:journals/pacmse/PizzornoB25, DBLP:journals/corr/abs-2601-09695}, and program repair~\cite{DBLP:conf/icse/XiaWZ23, DBLP:conf/nips/YangJWLYNP24, DBLP:conf/issta/0002RFR24}. In each case, a developer specifies the target behavior in natural language and an LLM produces code, tests, or patches from that specification. 
AI-augmented developer environments such as Cursor~\cite{anysphere2024cursor}, Claude Code~\cite{anthropic2025claude}, and GitHub Copilot~\cite{github2021copilot} have further accelerated this trend. As a result, developers increasingly operate at the level of natural language specification rather than code, a practice recently termed \emph{vibe coding}~\cite{Karpathy25}. 

The success of LLMs across these tasks suggests that they encode rich semantic knowledge about both code and natural language, and crucially, about the relationship between the two. This knowledge is reflected in the probability distributions LLMs assign to code sequences: Sengamedu and Zhao~\cite{DBLP:conf/sigsoft/SengameduZ22} show that LLMs assign lower probability to unnatural or incorrect code, and Palomba et al.~\cite{DBLP:journals/corr/abs-2401-12714} confirm that cross-entropy correlates with code maintainability. Most closely related to our work, Taherkhani et al.~\cite{DBLP:journals/tse/TaherkhaniSTMGH26} demonstrate that token-level entropy reliably distinguishes valid from invalid LLM-generated test cases. Taken together, these results suggest that LLM entropy is a meaningful signal for assessing the alignment between a code artifact and its intended specification, an insight we exploit when defining prompt coverage.

\subsection{Requirement-based Testing} 
Software requirements define the intended behavior of a system and serve as the foundation for both its development and validation~\cite{sommerville2011software}. Requirement-based testing formalizes this connection by deriving test cases directly from requirements, ensuring that the test suite validates whether the software implements the specified behavior. Early work established systematic procedures for deriving tests from structured specifications, including equivalence partitioning and boundary value analysis~\cite{myers2004art}, as well as model-based approaches that generate tests from UML state machines or activity diagrams~\cite{DBLP:conf/fm/Pretschner05}. A central challenge in requirement-based testing is traceability~\cite{DBLP:journals/tse/AntoniolCCLM02}: establishing and maintaining links between requirements and the test cases that exercise them. 

Traceability has been studied extensively in the context of structured requirements, using information retrieval~\cite{DBLP:conf/refsq/MertenKMSBP16}, machine learning~\cite{DBLP:conf/icse/0004CC17}, and more recently LLMs~\cite{DBLP:conf/icse/Fuchbeta0KLETK25} to recover or predict requirement--test links. However, these approaches assume that requirements and tests share a common vocabulary or that the relationship between them can be learned from labeled examples. When requirements take the form of informal natural language prompts, as in LLM-based code generation, neither assumption holds, and traceability becomes a semantic alignment problem rather than a lexical one.

\subsection{Coverage criteria} 
Test adequacy criteria provide a formal basis for assessing whether a test suite exercises a program sufficiently, guiding testers towards potential blind spots~\cite{myers2004art}. Classical structural criteria such as statement, branch, and condition coverage measure the proportion of code elements exercised by the test suite~\cite{ZhuHM97}. Specification-based criteria derive coverage targets from a formal behavioural description rather than from source code: examples include logic coverage criteria for LTL or CTL formulae~\cite{DBLP:conf/qsic/FraserA08} and transition coverage for finite state machines~\cite{DBLP:journals/stvr/UttingPL12}. 

However, all of these criteria either operate on the implementation (code) or require a formal abstraction of the system under test. In an era where developers operate at the level of natural language prompts, and where minor changes to a prompt can imply significant changes to the system under test, coverage should be defined at the specification level rather than the code level. To fill this gap, we introduce \emph{prompt coverage}, the first test adequacy criterion defined at the level of the natural language prompt, measuring which requirements are semantically aligned with test cases in the suite by exploiting the semantic code-related knowledge of LLMs.

%% file: sections/conclusion.tex
\section{Conclusion}

Large Language Models are rapidly reshaping software development by enabling the automatic synthesis of programs from natural language specifications. However, traditional test adequacy criteria rely on structural properties of source code rather than the requirements expressed in natural language prompts. This mismatch raises the question of whether traditional coverage metrics are sufficient for evaluating test suites in LLM-driven development settings. 

We introduced Prompt Coverage Adequacy, a novel testing criterion that measures the extent to which a test suite explores the semantic content of a prompt. Our approach leverages the attention mechanisms of LLMs to assess the relationship between candidate test cases and natural language requirements, using changes in model entropy as an indicator of prompt coverage. Unlike traditional coverage metrics, Prompt Coverage operates directly on the specification and does not require access to the implementation.

We evaluated Prompt Coverage on two widely used benchmarks, HumanEval and LiveCodeBench, using Qwen2.5 and Gemma 3 as the underlying models. The results demonstrate that prompt coverage is a meaningful and informative adequacy metric. Despite being entirely natural language-based, Prompt Coverage exhibits a strong positive relationship with code coverage, indicating that tests that better explore the prompt also tend to exercise larger portions of the generated implementation. Moreover, we showed that increased prompt coverage consistently leads to improved fault detection effectiveness, even when controlling for test suite size. This suggests that the observed gains stem from the quality and diversity of prompt exploration rather than simply from generating additional test cases.

Perhaps most importantly, our experiments revealed that Prompt Coverage-guided test generation achieves 30\% higher fault detection than Code Coverage-guided generation. This result highlights that prompt coverage captures aspects of program behavior that are not adequately reflected by structural implementation metrics alone. Consequently, prompt coverage should not be viewed merely as a surrogate for code coverage, but rather as a complementary criterion that provides a specification-based perspective on test adequacy.

More broadly, our findings suggest that Prompt Coverage can serve as a foundation for developing testing methodologies better aligned with the emerging paradigm of LLM-driven software engineering. As prompts increasingly play the role of executable specifications and implementations are generated automatically, adequacy criteria must evolve beyond exclusively code-based notions of coverage. By quantifying how thoroughly tests exercise the requirements encoded in prompts, Prompt Coverage provides a practical step toward specification-driven testing frameworks that remain effective even when source code is unavailable, highly variable, or continuously regenerated.

Moreover, user-in-the-loop testing, in which developers define and assess expected behaviors, e.g., definign test oracles, requires operating at a level of abstraction consistent with the development process. In a vibe-coding paradigm, where development is primarily driven by prompts rather than direct coding, manual user involvement should occur at the prompt level. Requiring developers to write and maintain tests at the code level would be misaligned with this workflow and could undermine the productivity benefits of prompt-based development.

Future work may explore richer representations of prompt semantics, extend Prompt Coverage to multi turn and agentic workflows, and investigate its applicability beyond code generation to other LLM-based tasks. We also envision combining prompt coverage with traditional code-based metrics to create hybrid adequacy criteria that capture both specification completeness and implementation thoroughness. Ultimately, we believe that prompt-aware testing metrics will play a central role in software quality assessment by ensuring the reliability, correctness, and trustworthiness of software produced through AI-assisted development.